\documentclass[pra,twocolumn,showpacs,eqsecnum]{revtex4}
\usepackage{graphicx}
\usepackage[urlcolor=blue, hyperindex, colorlinks, bookmarks=true,linkcolor=black,citecolor=black]{hyperref}
\begin{document}

\newcommand{\sch}{Schr\"odinger }
\newcommand{\schs}{Schr\"odinger's }
\newcommand{\nn}{\nonumber}
\newcommand{\nl}{\nn \\ &&}
\newcommand{\dg}{^\dagger}
\newcommand{\rt}[1]{\sqrt{#1}\,}
\newcommand{\bra}[1]{\langle{#1}|}
\newcommand{\ket}[1]{|{#1}\rangle}
\newcommand{\ito}{It\^o }
\newcommand{\str}{Stratonovich }
\newcommand{\erf}[1]{Eq.~(\ref{#1})}
\newcommand{\erfs}[2]{Eqs.~(\ref{#1}) and (\ref{#2})}
\newcommand{\erft}[2]{Eqs.~(\ref{#1}) -- (\ref{#2})}

\title{Jump-like unravelings for non-Markovian open quantum systems}
\date{\today}
\author{Jay Gambetta}
\affiliation{Centre for Quantum Dynamics, School of Science,
Griffith University, Brisbane 4111, Australia}
\author{T. Askerud}
\affiliation{Centre for Quantum Dynamics, School of Science,
Griffith University, Brisbane 4111, Australia}
\author{H. M. Wiseman} \email{h.wiseman@griffith.edu.au}
\affiliation{Centre for Quantum Dynamics, School of Science,
Griffith University, Brisbane 4111, Australia}

\begin{abstract}
Non-Markovian evolution of an open quantum system can be
`unraveled' into pure state trajectories generated by a
non-Markovian stochastic (diffusive) \sch equation, as introduced
by Di\'osi, Gisin, and Strunz. Recently we have shown that such
equations can be derived using the modal (hidden variable)
interpretation of quantum mechanics. In this paper we generalize
this theory to treat jump-like unravelings. To illustrate the
jump-like behavior we consider a simple system: A classically
driven (at Rabi frequency $\Omega$) two-level atom coupled
linearly to a three mode optical bath, with a central frequency
equal to the frequency of the atom, $\omega_0$, and the two side
bands have frequencies $\omega_0\pm\Omega$. In the large $\Omega$
limit we observed that the jump-like behavior is similar to that
observed in this system with a Markovian (broad band) bath. This
is expected as in the Markovian limit the fluorescence spectrum
for a strongly driven two level atom takes the form of a Mollow
triplet. However the length of time for which the Markovian-like
behaviour persists depends upon {\em which} jump-like unraveling
is used.
\end{abstract}


\maketitle

\section{Introduction}

In the past few years non-Markovian stochastic \sch equations
(SSEs) for diffusive unravelings have received considerable
attention
\cite{DioStr97,DioGisStr98,StrDioGis99,YuDioGisStr99,Cre00,
Bud00,BasGhi02,GamWis002,GamWis003,Bas03,GamWis005,GamWis006}. A
diffusive non-Markovian stochastic \sch equation is a non-linear,
non-Markovian, and stochastic evolution equation for a quantum
state.  We usually denote this quantum state by
$\ket{\psi_{\{z_k(t)\}}(t)}$ where the subscripts $\{z_k(t)\}$
implies that $\ket{\psi_{\{z_k(t)\}}(t)}$ is conditioned on the
set of continuous time dependent random variables $\{z_k(t)\}$
described by a probability distribution $P(\{z_k\},t)$. This set
is constrained such that
\begin{equation}\label{eq.RhoFromNonMarkovianSSE}
  \rho_{\rm red}(t)=E[\ket{\psi_{\{z_k(t)\}}(t)}\bra{\psi_{\{z_k(t)\}}(t)}],
\end{equation} where $E[...]$ denotes an ensemble average over the
set $\{z_k(t)\}$. Here $\rho_{\rm red}(t)$ refers to the reduced
state of an open quantum system. An open quantum system is a
combined system consisting of a ``system of interest'' and an
environment or bath. The reduced state is defined by
\begin{equation}\label{eq.RhoFromTrace}
  \rho_{\rm red}(t)={\rm Tr}_{\rm
  env}[\ket{\Psi(t)}\bra{\Psi(t)}].
\end{equation} Here $\ket{\Psi(t)}$ is the solution of the \sch
equation for the combined system (generally an entangled state)
and ${\rm Tr}_{\rm
  env}[...]$ refers to a partial trace performed over the
environments degrees of freedom.

In this paper we extend the known unravelings for non-Markovian
SSEs to include two jump-like unravelings. This is motivated by
the fact that in the Markovian limit the simplest of all Markovian
SSEs is jump-like in nature. This corresponding to the direct
detection quantum trajectory \cite{Car93}. However, as shown in
Ref. \cite{GamWis002} in the non-Markovian case SSEs are {\em not}
quantum trajectories. That is, they are not evolution equations
for the state of the system conditioned on the outcomes of some
continuous-in-time measurements of the bath. This is because a
non-Markovian system remembers what has happen in the past
(whether a measurement has occurred or not) and a continuously
measured system will have a drastically different average
evolution, $\rho(t)$, from a non-measured system $\rho_{\rm
red}(t)$.

So what is the interpretation of non-Markovian SSEs? In Ref.
\cite{GamWis005} we came to the conclusion that the three current
non-Markovian SSEs, the coherent
\cite{DioGisStr98,Cre00,GamWis002}, quadrature
\cite{BasGhi02,GamWis002} and the position \cite{GamWis006} only
have a non-trivial (non-numerical) interpretation when we consider
the modal
\cite{Fra81,Hea89,Die97,Bub97,BacDic99,Sud00,SpeSip01b,GamWis004}
(or Bell's beable \cite{Bel84}) interpretation of quantum
mechanics. They are evolution equations for the system part of the
property state of the universe when certain bath observables are
given an objective reality (become hidden variables). The values
of these bath hidden variables, at time $t$, are given by the set
$\{z_k(t)\}$. The different unravelings correspond to a different
choice of which bath observables are to be objective real. For
example the position unraveling arises when the hidden variables
for the bath correspond to the positions of each oscillator making
up the bath \cite{GamWis005,GamWis006}. That is, the set
$\{z_k(t)\}$ becomes the set $\{x_k(t)\}$ corresponding to the
values of the positions of each oscillator of the bath. For this
case the position trajectories, $\{d_t x_k(t)\}$, actually
correspond to Bohmian trajectories \cite{Boh52a}.

Note this is not the only interpretation of non-Markovian SSEs; in
Refs. \cite{BasGhi02} and \cite{BasGhi03} Bassi and Ghirardi
interpreted these equations as a new dynamical reduction modal,
dynamical reduction with non-white Gaussian noise. In this paper
we are not going to argue against this interpretation except to
say that by accepting the above modal interpretation we can easily
generalize the theory for diffusive non-Markovian SSEs to include
jump-like ``unravelings.''

A difference from the diffusive case is that it does not appear
possible to derive an explicit expression for the jump-like
non-Markovian SSEs. This is because these non-Markovian SSEs are
conditioned on a discrete set $\{z_{n_k}(t)\}$ and as such it is
impossible to define derivatives with respect to this set.
Furthermore a Girsanov transformation \cite{DioGisStr98} can not
be defined. In fact it turns out that to derive numerical
jump-like non-Markovian unravelings is effectively the same as
solving the \sch equation for the combined system. Thus any
numerical advantage of using non-Markovian SSE to determine
$\rho_{\rm red}(t)$ [see Eq. (\ref{eq.RhoFromNonMarkovianSSE})] is
lost for these unravelings. However, the point of this paper is
not to derive more effective ways of finding $\rho_{\rm red}(t)$
but rather to illustrate that by using the modal interpretation we
can derive the numerical equivalent of the solution to a
non-Markovian SSE for jump-like unravelings.

To illustrate a non-Markovian unraveling that exhibit jump-like
behavior we consider the two unravelings, spectral-mode and
temporal-mode. We apply our theory to a simple system: A
classically driven two level atom (TLA), at Rabi frequency
$\Omega$, coupled linearly to a three-mode optical bath, with a
central frequency equal to the frequency of the atom,
$\omega_{0}$, and the two side bands have frequencies
$\omega_{0}\pm\Omega$. In the large $\Omega$ limit we observed
that the jump-like behavior is similar to that observed in
Markovian systems. This is expected as in the Markovian limit a
strongly driven TLA has a fluorescent spectrum which takes the
form of a Mollow triplet \cite{Mol75} with peaks at these
frequencies.

The structure of this paper is as follows: In the next section we
present an outline of the general physical model to which the
theory of non-Markovian SSE is applicable. In  Sec.
\ref{Sec.Modal} we present a review of the modal interpretation of
quantum mechanics and how it applies to non-Markovian SSEs. In
Secs. \ref{Sec.number} and \ref{Sec.XXX} we derive two jump-like
unravelings, these being the spectral-mode and the temporal-mode
unraveling. In Sec \ref{Sec.Coll} we briefly outline how Jack,
Collet and Walls \cite{JacColWal99,JacColWal99b,JacCol00}
non-Markovian quantum trajectory theory differs to our theory.
Note other jump-like non-Markovian unravelings exists
\cite{BreKapPet99,Bre03}, but these are only numerical methods for
solving $\rho_{\rm red}(t)$, and as such will not be considered in
this paper. Finally in Sec. \ref{Sec.Conclude} we conclude this
paper.

\section{The TLA and the underlying bath dynamics}\label{Sec.Dynamics}

The aim of this section is to outline the underlying model we will
be using to generate jump-like non-Markovian unravelings. The
standard model is to assume a preferred factorization of the
universe into a system and environment, label by the Hilbert
spaces ${\cal H}_{\rm sys}$ and ${\cal H}_{\rm env}$ respectively.
The total Hamiltonian for the universe is then given by
\begin{equation}\label{eq.Hamiltonian}
 \hat{H}_{\rm sys}(t)\otimes\hat{1}+\hat{1}\otimes\hat{H}_{\rm
 env}+\hat{V}.
\end{equation} Here $\hat{H}_{\rm sys}$, $\hat{H}_{\rm env}$ and
$\hat{V}$ represent the Hamiltonians for the system, environment
and any interaction taking place between the system and
environment respectively. In this paper we assume the system is a
two level atom (TLA) described by the free system Hamiltonian
\begin{equation}\label{eq.SystemHamiltonian}
  \hat{H}_{\rm sys}(t)=\hat{H}_{\rm sys_{0}}+\hat{H}(t)=\frac{\hbar \omega_0}{2}\hat{\sigma}_{z}+\hat{H}(t),
\end{equation} where $\omega_0$ is the atomic transition frequency
and $\hat{\sigma}_{z}$ is the Pauli spin operator
\begin{equation}\label{eq.PauliSpinZ}
\hat{\sigma}_z=\ket{e}\bra{e}-\ket{b}\bra{b},
\end{equation}
where $\ket{e}$ and $\ket{b}$ represent the excited and ground
state of the atom. The extra term $\hat{H}(t)$ in Eq.
(\ref{eq.SystemHamiltonian}) represents any extra system
evolution, for example a classical driving process.

The environment, as in all optical situations, is modelled by a
collection of one-dimensional harmonic oscillators. The free
Hamiltonian for this type of environment is
\begin{equation}\label{eq.EnvironmentHamiltonian}
  \hat{H}_{\rm env}=\sum_{k}^{\kappa}{\hbar \omega_k}\hat{a}\dg_k\hat{a}_k,
\end{equation} where $\kappa$ defines the total number of modes, while $\omega_k$
  and $\hat{a}_k$ ($\hat{a}_k\dg$) are
the frequency and the annihilation (creation) operator of the
$k^{\rm th}$ mode of the environment.

We assume that the interaction between the system and environment
is consistent with the rotating wave approximation. That is, we
assume $\hat{V}$ is linear in the environment amplitudes, and has
the form
\begin{equation}\label{eq.InteractionHamiltonian}
\hat{V}=i\hbar\sum_{k}^{\kappa}(g_k^{*}\hat{\sigma}\hat{a}_k
\dg-g_k\hat{\sigma}\dg\hat{a}_k),
\end{equation} where $g_k$ is the coupling strength of the $k^{\rm
th}$ mode to the system.

For calculational purposes we define an interaction frame such
that the Hamiltonians $\hat{H}_{\rm sys_{0}}$ [which is defined
implicity in Eq. (\ref{eq.SystemHamiltonian})] and $\hat{H}_{\rm
env}$ is removed. The unitary evolution operator for this
transformations is
\begin{equation}\label{eq.UnitaryFree}
\hat{U}_{0}(t,0)=e^{-i(\hat{H}_{\rm sys_{0}}\otimes\hat{1}+\hat{1}
\otimes\hat{H}_{\rm env})(t-0)/{\hbar}}.
\end{equation} Thus the combined state in the interaction frame is defined as
\begin{equation}\label{eq.UnitarySch} \ket{\Psi(t)}=\hat{U}_{0}\dg(t,0)\ket{\Psi(t)_{\rm Sch}},
\end{equation} and an arbitrary operator $\hat{A}$ becomes \begin{equation} \hat{A}_{\rm
int}(t)=\hat{U}_{0}\dg(t,0)\hat{A}\hat{U}_{0}(t,0). \end{equation}
This allows
 us to write the \sch equation as
\begin{equation} \label{eq.IntSchEquation} d_t\ket{{\Psi}(t)} =-\frac{i}{\hbar}[\hat{H}_{\rm int}(t)+\hat{V}_{\rm int}(t)]
\ket{\Psi(t)},
\end{equation} where $\hat{H}_{\rm int}(t)$ refers to $\hat{H}(t)$ in
the interaction frame and $\hat{V}_{\rm int}(t)$ is defined as
\begin{eqnarray} \label{eq.HamiltonianInteractionInt} \hat{V}_{\rm
int}(t)= i\hbar\sum_{k}^{\kappa}[\hat{\sigma}g^*_{k}e^{i\Omega_{k}
t}\hat{a}_{k}\dg-\hat{\sigma}\dg g_{k}e^{-i\Omega_{k}
t}\hat{a}_{k} ],
\end{eqnarray} where $\Omega_{k}=\omega_{k}-\omega_0$.

In all the examples we present in this paper we take the extra
system Hamiltonian, $\hat{H}(t)$, to be classical driving of the
TLA. Under the dipole and rotating wave approximation the extra
system Hamiltonian in Eq. (\ref{eq.SystemHamiltonian}) is
\begin{equation}\label{eq.Hsch}
  \hat{H}(t)=\frac{\hbar\Omega}{2}[\hat{\sigma}\dg\exp(-i\omega_{\rm
  c}t)+\hat{\sigma}\exp(i\omega_{\rm c}t)],
\end{equation}
where $\Omega=-2 {\bf d}\cdot{\bf E}_{0}({\bf r}_0)/\hbar$ is the
Rabi frequency and ${\bf E}_{0}({\bf r}_0)$ is the amplitude of
the classical field at the positions of the atom and ${\bf d}$ is
dipole transition matrix.  For simplicity we tune the frequency of
the classical driving field, $\omega_{\rm c}$, to the atomic
transition frequency. Thus when moving to the interaction frame
Eq. (\ref{eq.Hsch}) becomes
\begin{equation}\label{eq.Hint}
  \hat{H}_{\rm
  int}=\frac{\hbar\Omega}{2}(\hat{\sigma}\dg+\hat{\sigma})=\frac{\hbar\Omega}{2}\hat{\sigma}_x,
\end{equation} which is time independent.

\section{Modal interpretation of quantum mechanics}\label{Sec.Modal}

\subsection{General modal dynamics}

In this section we give a brief overview of the modal
interpretation of quantum mechanics; for a more detailed
description see Refs.
\cite{Bel84,Bub97,BacDic99,Sud00,SpeSip01b,GamWis004}. The basic
idea of this view of quantum mechanics is that certain observables
have an objective reality independent of measurement. This is in
contrast to the orthodox interpretation where reality is undefined
prior to measurement and it is the act of observation that defines
the reality of the observable.

In quantum mechanics it is convenient to define an observable as
\begin{equation}\label{eq.Obser}
   Z=\{(z_{n},\hat{\pi}_{n})\},
\end{equation} where the set of values $\{z_{n}\}$ correspond to the possible values the observable can
have and $\{\hat{\pi}_{n}\}$ is the projective measure for the
observable. When these values correspond to numbers we can write
an observable as an operator
\begin{equation}\label{eq.ObservableMatrix}
  \hat{Z}=\sum_{n}z_n\hat{\pi}_n.
\end{equation} The projectors are orthogonal
and form a decomposition of unity:
\begin{equation}
 \sum_n \hat\pi_n = \hat{1}.
\end{equation}

In quantum mechanics because of the non-commutative nature of
observables, not all observables can be written in terms of the
same projective measure $\{\hat{\pi}_{n}\}$. In the modal theory
one postulates that only one set of projectors has an objective
reality, this being called the preferred projective measure. Once
this measure is specified it uniquely defines a group of
observables which are objectively real, this being the group
defined to have elements given by Eq. (\ref{eq.Obser}). In this
paper we will refer to this group as the properties of the system.
That is, the group of preferred observables (observables with
objective reality) are labelled properties of the system.

The biggest criticism against the modal theory is that many
choices of the preferred measure are viable \cite{Bub97}. This has
lead to many variants of the modal interpretation of quantum
mechanics
\cite{Bel84,Bub97,BacDic99,Sud00,GamWis004,Boh52a,Hea89,Die97,SpeSip01b}.
Some have tried to address the problem of choice, either by
accepting it (the beable variant) \cite{Bel84,Bub97,Sud00},
selecting the position projective measure as preferred (Bohmian
mechanics \cite{Boh52a}), or by applying an algorithm for
determining the preferred projective measure based on the total
wavefunction of the universe
\cite{Bub97,BacDic99,Hea89,Die97,SpeSip01b}. The algorithmic
approaches in our opinion still face choice as one has to choose
the preferred factorization of the universe. Furthermore we have
shown in Ref. \cite{GamWis004} that we can extend the modal theory
to include positive operator measures, thereby increasing the
amount of possible choices. Here, however, we will only consider
preferred projective measures and take the view that choice is
fundamental. Depending on the physical situation we wish to
describe we will choose the appropriate measure $\{\hat\pi_n\}$.

Once we have chosen the preferred projective measure, to explain
why property $Z$ has the value $z_n$ at time $t$ we can introduce
an extra quantum state, the {\em property state}. It is defined as
\begin{equation}\label{eq.propertystate}
\ket{\Psi_{z_n(t)}(t)}=\hat\pi_{n}\ket{\Psi(t)}/\sqrt{N},
\end{equation} where $N$ is a normalization constant and $z_n(t)$ is the value of
$Z$ at time $t$. The property state at time $t$ is determined by
both $\ket{\Psi(t)}$ and a stochastic evolution (jumps between
different $n$) .
 It is interpreted as the
actual state of the universe, and by the eigenstate-eigenvalue
link it selects the present value $z_n$ for property $Z$ from the
possible values $\{z_n\}$. The stochastic dynamics (rates at which
it jumps between different $n$) is determined by $\ket{\Psi(t)}$
and as such $\ket{\Psi(t)}$ in this interpretation is called the
guiding state.

The modal dynamics (the stochastic evolution of the property
state) is found using the method originally proposed by Bell
\cite{Bel84} and generalized in Refs.
\cite{BacDic99,Sud00,GamWis004}. We start by defining ${\rm
Pr}(z_n,t)$ as the probability that the property will have the
value $z_n$ at time $t$. Assuming a Markovian process, by which we
mean that the probability of the property being $z_m$ at time
$t+dt$ only depends on the value at time $t$, we can write a
master equation for ${\rm Pr}(z_n,t)$ as
\begin{equation}\label{eq.Master2}
  d_t {\rm Pr}(z_n,t)=\sum_{m}[T_{nm}(t){\rm Pr}(z_m,t)-T_{mn}{\rm Pr}(z_n,t)],
\end{equation} where $T_{nm}$ (for $n\neq m$) are
transition rates. For $n=m$, $T_{nn}$ (which is negative) is a
measure of the rate at which value $z_n$ loses probability.

Defining a probability current $J_{nm}(t)$ as
\begin{equation}\label{eq.probcurrent}
J_{nm}(t)=T_{nm}(t){\rm Pr}(z_m,t)-T_{mn}{\rm Pr}(z_n,t),
\end{equation} results in $J_{nm}(t)=-J_{mn}(t)$ and
allows us to rewrite the probability master equation as
\begin{equation} \label{eq.master3}
d_t {\rm Pr}(z_n,t)=\sum_{m}J_{nm}(t).
\end{equation} Given $J_{nm}(t)$ and ${\rm Pr}(z_n,t)$, there
are many possible transition rates satisfying \erf{eq.master3}.
One solution, chosen by Bell \cite{Bel84} is as follows.

For $J_{nm}(t)<0$,
\begin{eqnarray}\label{eq.t1}
  T_{nm}(t)&=&0, \\
  T_{mn}(t)&=&-J_{nm}(t)/{\rm Pr}(z_n,t),
\end{eqnarray} and for $J_{nm}(t)>0$
\begin{eqnarray}
  T_{nm}(t)&=&J_{nm}(t)/{\rm Pr}(z_m,t),\\
  T_{mn}(t)&=&0. \label{eq.t4}
  \end{eqnarray}
Thus once we have the probability current it is possible to
calculate the transition matrix $T_{nm}(t)$ which in turn allows
us to calculate (via using a random number generator) a trajectory
for the value of the property $Z$.

To find $J_{nm}(t)$, as in the orthodox theory, we have to
postulate a fundamental rule for probability,  the Born rule
\begin{equation}\label{eq.BornRule}
  {\rm Pr}(z_n,t)=\bra{\Psi(t)}\hat{\pi}_n\ket{\Psi(t)}.
\end{equation} With this equation and \erfs{eq.master3}{eq.IntSchEquation}
a possible solution for $J_{nm}(t)$ is
\begin{eqnarray} \label{eq.BellJ}
  J_{nm}(t)&=&2{\rm Im}\{\bra{\Psi(t)}\hat{\pi}_{n}[\hat{H}_{\rm int
}(t)+\hat{V}_{\rm
int}(t)]\nl\times\hat{\pi}_{m}\ket{\Psi(t)}\}{/\hbar}.
\end{eqnarray}
Note the ensemble set of trajectories we obtained for the value of
the property $Z$ is only one of the infinitely many possible sets
of ensemble trajectories. Others can be found by either adding an
extra term $T^{0}_{nm}(t)$ to $T_{nm}(t)$, where $T^{0}_{nm}(t)$
is constrained only by
\begin{equation}
T_{nm}^{0}(t)P_{m}(t)-T_{mn}^{0}(t)P_{n}(t)=0,
\end{equation} or by adding any current $J_{nm}^{~0}(t)$ to
$J_{nm}(t)$ which satisfies
\begin{equation}
\sum_{m} J_{nm}^{~0}=0.
\end{equation}
 For the purposes of this paper we only consider Bell solution
[not containing the extra $T_{nm}^{0}(t)$ and $J_{nm}^{~0}(t)$
terms]. For a discussion of these solutions see Refs. \cite{Vin93}
or \cite{Dic97}.

\subsection{Application to non-Markovian SSEs}

In Ref. \cite{GamWis005} we showed that the theory of the
non-Markovian SSEs emerge from modal dynamics when we assume a
preferred projective measure of the form
\begin{equation}
\{ \hat{\pi}_{n}=\hat{1}_{\rm sys}\otimes \hat{\pi}_{n_{\rm
env}}\},
\end{equation} where $\hat{\pi}_{n_{\rm env}}$ is a projector
define solely to operate in the Hilbert space of the environment.
(Note the coherent non-Markovian SSE
\cite{DioStr97,DioGisStr98,StrDioGis99,YuDioGisStr99,Cre00,
GamWis002,GamWis005} arises when we use a preferred POM rather
than a projective measure; see Ref. \cite{GamWis005}). This means
the observables which have definite values are of the form
\begin{equation}
  Z=\{z_n, \hat{1}_{\rm sys}\otimes \hat{\pi}_{n_{\rm
env}}\}.
\end{equation} That is, the bath is given property status,
while the system is treated as a purely quantum system, which
nevertheless influences the bath values via the coupling
Hamiltonian [Eq. (\ref{eq.HamiltonianInteractionInt})].

Diffusive non-Markovian arise when the projective measure for the
environment $\{\hat{\pi}_{n_{\rm env}}\}$ is chosen to be an
infinitesimal projective density measure $\{\hat{\pi}(z)_{\rm
env}dz\}$. That is, properties are given by
\begin{equation}
  Z=\{z, \hat{1}_{\rm sys}\otimes \hat{\pi}(z)_{\rm
env}dz\},
\end{equation} where $Z$ now has a continuous set of possible values $\{z\}$.
For example the position unraveling occurs when we choose the
following infinitesimal projective density measure
\begin{equation}
  \{\hat{\pi}(x_1)dx_1\otimes ...\otimes
\hat{\pi}(x_k)dx_k\otimes...\otimes
\hat{\pi}(x_\kappa)dx_\kappa\},
\end{equation} where $\hat{\pi}(x_k)=\ket{x_k}\bra{x_k}$ and
$\ket{x_k}$ is defined to be the eigenstate of the operator
$\hat{x}_k=(\hat{a}_k+\hat{a}_k\dg)/\rt{2}$. That is, the
observables correspond to the positions of the environment's
$\kappa$ harmonic oscillators (and any function of them by the
principle of property compositions \cite{BacDic99}) are given
property status. With this choice of preferred projective measure
the property state of the universe becomes
\begin{equation}\label{eq.propertystate2}
\ket{\Psi_{\{x_k(t)\}}(t)}=\ket{\{x_k(t)\}}\otimes\ket{\psi_{\{x_k(t)\}}(t)},
\end{equation} where $\ket{\psi_{\{x_k\}}(t)}$ is called the
conditional system state. Here we see that the property state is
divided into two parts, an environment state and system state.

In Ref. \cite{GamWis005} we showed that when using Bell's solution
for the transition parameters we can derive a set of $\kappa$
trajectories for the values of the $\kappa$ position properties,
which we denote as $\{x_k(t)\}$ (these actually turn out to be
Bohmian trajectories). Using these trajectories we can derive a
differential equation for $\ket{\psi_{\{x_k(t)\}}(t)}$, which
describes the evolution of the system part of the property state
of the universe, the environment part is given by
$\ket{\{x_k(t)\}}$. This system-state differential equation turns
out to be the non-Markovian SSE for the position unraveling. For
the complete derivation of this non-Markovian SSE and the other
two diffusive non-Markovian SSEs see Refs.
\cite{DioGisStr98,GamWis002,GamWis005,GamWis006}.

Here we are going to use the above reasoning to motivate how one
can go about finding numerically the equivalent to the conditioned
system state for jump-like unravelings.  To do this we simply
choose discrete rank one projectors for the environment.  That is,
the preferred projective measure is
\begin{equation}
   \{\hat{\pi}_{n}\}= \{\hat{1}_{\rm sys}\otimes\ket{\{z_{n_k}\}}\bra{\{z_{n_k}\}}\},
\end{equation} where
$\ket{\{z_{n_k}\}}=\ket{z_{n_1}}\otimes...\otimes\ket{z_{n_k}}\otimes
...\otimes\ket{z_{n_\kappa}}$ and $n=\{n_1,n_2,...,n_\kappa\}$.
With this preferred projective measure the property state of the
universe [see Eq. (\ref{eq.propertystate})] becomes
\begin{equation}\label{eq.propertystate3}
\ket{\Psi_{\{z_{n_k}(t)\}}(t)}=\ket{\{z_{n_k}(t)\}}\otimes\ket{\psi_{\{z_{n
_k}(t)\}}(t)},
\end{equation}
where the conditioned state is
\begin{equation}\label{eq.PropertySystemState}
\ket{\psi_{\{z_{n
_k}(t)\}}(t)}=\langle\{z_{n_k}(t)\}\ket{\Psi(t)}/\sqrt{N}.
\end{equation} The evolution of this state through time is equivalent to a
non-Markovian SSE for jump-like unravelings.

To summarize, the procedure to developed numerical trajectories
for both $\ket{\psi_{\{z_{n _k}(t)\}}(t)}$ and $\{z_{n_k}(t)\}$ is
to first calculate the guiding state from the \sch equation [see
Eq. (\ref{eq.IntSchEquation})], then using Bells solution we can
determine both $J_{nm}(t)$ and $T_{nm}(t)$ for the appropriately
chosen preferred projective measure. Once these are calculated we
use a random number generator to simulate simultaneously a typical
trajectory for $\{z_{n_k}(t)\}$ and hence $\ket{\psi_{\{z_{n
_k}(t)\}}(t)}$.

\section{The spectral mode unraveling}\label{Sec.number}

The first unraveling we consider is the simplest of all; it
corresponds to the bath hidden variables being the photon number
of each bath spectral mode. That is, the values of the bath hidden
variables are denoted by $\{n_{k}\}$ and only take on discrete
values. The preferred projective measure for this unraveling is
\begin{equation}
   \{\hat{\pi}_{n}\}= \{\hat{1}_{\rm sys}\otimes\ket{\{{n_k}\}}\bra{\{{n_k}\}}\},
\end{equation} where
$\ket{\{{n_k}\}}=\ket{{n_1}}\otimes...\otimes\ket{{n_k}}\otimes
...\otimes\ket{{n_\kappa}}$ and $\ket{{n_k}}$ is the eigenstate of
the spectral number operator $\hat{n}_k=\hat{a}_k\dg\hat{a}_k$.

To illustrate this unraveling we consider two cases of a driven
TLA coupled to the bath. In the first the `bath' is a single mode
harmonic oscillator and in the second the bath consists of three
modes as described in the introduction.

\subsection{A single-mode bath}

The first example consists of a bath with only one mode
($\kappa=1$). Thus it only has one bath hidden variable with value
$n$ at time $t$. Thus $n(t)$ denotes the trajectory of this value
through time. By Eq. (\ref{eq.HamiltonianInteractionInt}) the
interaction Hamiltonian in the interaction frame with the free
dynamics removed is
\begin{eqnarray} \hat{V}_{\rm
int}(t)= i\hbar[g^*e^{i(\omega_1-\omega_0)t}
\hat{\sigma}\hat{a}\dg-ge^{-i(\omega_1-\omega_0 )t}\hat{\sigma}\dg
\hat{a}].
\end{eqnarray} Choosing the bath frequency, $\omega_1=\omega_0$ and including the driving of the TLA
the evolution for the guiding state is determined by
\begin{equation} \label{eq.IntSchEquationSimple} d_t\ket{{\Psi}(t)}
=\Big{[}-\frac{i\Omega}{2}\hat{\sigma}_x+g^*
\hat{\sigma}\hat{a}\dg-g\hat{\sigma}\dg \hat{a}\Big{ ]}
\ket{\Psi(t)}.
\end{equation}
To calculate the stochastic evolution of the property state,
$\ket{\Psi_{n(t)}(t)}=\ket{n(t)}\ket{\psi_{n(t)}(t)}$, we use
Bell's solution to find $J_{nm}(t)$ [see Eq. (\ref{eq.BellJ})].
Doing this we get
\begin{equation} \label{eq.BellJSimple}
  J_{nm}(t)=2{\rm Im}\{\bra{\tilde{\psi}_n(t)}\bra{n}[ig^*
\hat{\sigma}\hat{a}\dg-ig \hat{\sigma}\dg
\hat{a}]\ket{m}\ket{\tilde{\psi}_m(t)}\}.
\end{equation} where $\ket{\tilde{\psi}_n(t)}=\langle{n}\ket{\Psi(t)}$ is the unnormalised conditioned
state in configuration space. This simplifies to
\begin{eqnarray}\label{eq.BellJSimple2}
   J_{nm}(t)&=&-2{\rm
   Im}[ig\bra{\tilde{\psi}_n(t)}\hat{\sigma}\dg\ket{\tilde{\psi}_m(t)}\rt{n+1}\delta_{n+1,m}]\nl+2{\rm
   Im}[ig^*\bra{\tilde{\psi}_n(t)}\hat{\sigma}\ket{\tilde{\psi}_m(t)}\rt{n}\delta_{n-1,m}].{\hspace{.7cm}}
\end{eqnarray}
Here we see that for a given $m$ only $J_{m\pm 1,m}(t)$ are
non-zero. Thus the only transitions allowed are from $m\rightarrow
m\pm 1$. The transition matrices [see Eqs. (\ref{eq.t1}) -
(\ref{eq.t4})] are as follows.

For $J_{m+ 1,m}(t)>0$ (or $J_{m-1,m}(t)<0$),
\begin{eqnarray}\label{eq.tsimple1}
  T_{m+1,m}(t)&=&J_{m+1,m}(t)/{\rm Pr}(z_m,t)\nn\\
  &=&2{\rm
   Im}\Big{[}ig^*\bra{{\psi}_{m+1}(t)}\hat{\sigma}\ket{{\psi}_m(t)}\nl\times\rt{\frac{({m+1}){\rm Pr}({m+1},t)}{{\rm Pr}(m,t)
   }}\Big{]},\\
  T_{m-1,m}(t)&=&0,
\end{eqnarray} and for $J_{m+
1,m}(t)<0$ (or $J_{m-1,m}(t)>0$),
\begin{eqnarray}
  T_{m+1,m}(t)&=&0, \\\label{eq.tsimple2}
  T_{m-1,m}(t)&=&J_{m-1,m}(t)/{\rm Pr}(z_m,t)\nn\\
  &=&-2{\rm
   Im}\Big{[}ig\bra{{\psi}_{m-1}(t)}\hat{\sigma}\dg\ket{{\psi}_m(t)}\nl\times
   \rt{\frac{m{\rm Pr}({m-1},t)}{{\rm Pr}(m,t)}}\Big{]} .
\end{eqnarray} Using a random number generator we can simulate a
typical trajectory for $n(t)$ based on the above transition rates.
For example given that $n(t)$ is $m$ at time $t$, depending on
which one of $J_{m\pm1,m}(t)$ is positive [$J_{m+1,m}(t)$ or
$J_{m-1,m}(t)$] determines whether the value of the bath hidden
variable is going to jump up or down in the interval $dt$. Let us
assume $J_{m+1,m}(t)$ is positive then the probability of an
upward jump in the interval $dt$ is $T_{m+1,m}(t)dt$, the
simulation works by choosing a random variable between $[0,1]$ and
if $T_{m+1,m}(t)dt$ is greater than this an upward jump occurs.
This procedure is then repeated until the desired simulation time
is reached. Note in \erft{eq.tsimple1}{eq.tsimple2} how the
conditioned state in configuration space, $\ket{\psi_n(t)}$,
guides $n(t)$. However, unlike diffusive non-Markovian we can see
here, explicitly, that it is impossible to define this transition
in terms of only one conditioned state, thus no analytical
evolution equation for $\ket{\psi_{n(t)}(t)}$ can be derived.

\begin{figure}
\includegraphics[width= .45\textwidth]{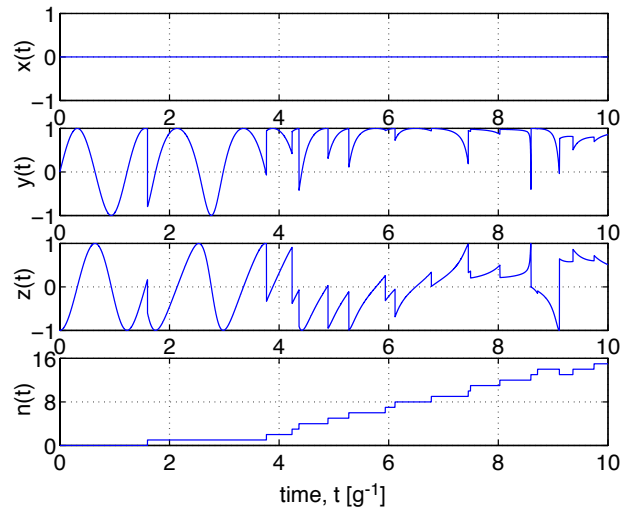}
\caption{\label{Fig.SingleModeNumber} This figure depicts a
typical trajectory in Bloch representation for the spectral mode
unraveling when the system is a driven TLA and the bath is a
single mode harmonic oscillator. In this figure all calculations
were done using the initial system state $\ket{\psi(0)}=\ket{b}$
with system parameters $g=1$, $\Omega=5$. Time is measured in
units $g^{-1}$.}
\end{figure}

Choosing $\Omega=5g$ a numerical simulation for a typical
trajectory is shown in Fig. \ref{Fig.SingleModeNumber} in Bloch
representation
\begin{eqnarray}\label{Bloch}
  x(t)&=&\langle\hat{\sigma}_x\rangle_t, \\
y(t)&=&\langle\hat{\sigma}_y\rangle_t,\\
z(t)&=&\langle\hat{\sigma}_z\rangle_t,
\end{eqnarray} where $\langle\hat{A}\rangle_t=\bra{\psi_{n(t)}(t)}\hat{A}\ket{\psi_{n(t)}(t)}$.
Also shown in this figure is the trajectory for the value of the
hidden value, $n(t)$. Here it is observed that the conditioned
state evolves smoothly until there is a jump in $n(t)$. Most of
jumps are up in photon number (all except one which occurs at
large $t$). This is because with classical driving we are
effectively adding energy to the system and for it to dissipate
this energy photons are emitted from the atom into the bath. This
is further verified by the fact that for most jumps there is a
corresponding lowering of atomic energy [that is, a lowering of
$z(t)$]. Note for Markovian dynamics an upward jump in photon
number always puts the system in the ground state ($z=-1$).
However in this figure this is not observed and in fact (for large
time in the figure) there are upward jumps in atomic energy when
the value of $n(t)$ increases. This clearly shows that
non-Markovian dynamics is a lot more complicated than Markovian.
Note the above strange behaviour (not Markovian-like) becomes more
evident at later times, when there are many photons in the bath.
This is expected since in the Markovian case there is never more
then one photon per bath mode on average.

\begin{figure}
\includegraphics[width= .45\textwidth]{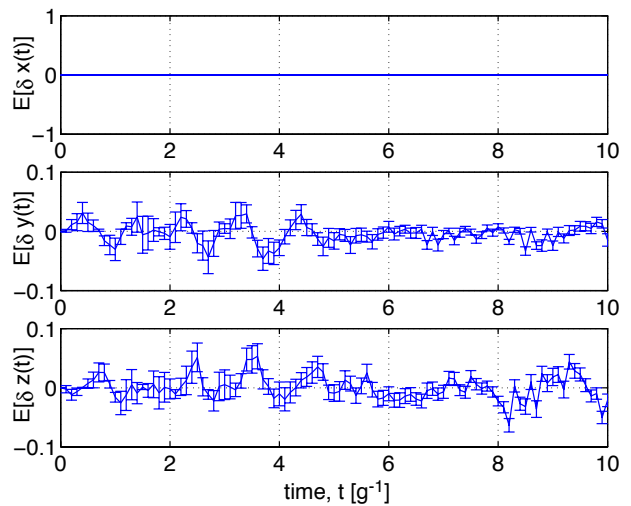}
\caption{\label{fig.Diff} This figure depicts the difference
between the ensemble average of 1000 trajectories and the actual
reduced state for the spectral mode unraveling when the system is
a driven TLA and the bath is a single mode harmonic oscillator.
All parameters are defined in Fig. \ref{Fig.SingleModeNumber}. }
\end{figure}

To show that by performing an ensemble average over these
trajectories do in fact give the reduced state, the difference
between the ensemble average of 1000 trajectories and the actual
reduced state is shown in Fig. \ref{fig.Diff}. Here it is observed
that to within statistical error these methods agree.

\subsection{The three mode bath}

In the above example we observed that it is possible to derive
numerically the equivalent to a non-Markovian SSE for the spectral
mode unraveling for a simple system. Now we consider a slightly
more complicated system: a TLA coupled linearly to a three mode
optical bath, with a central frequency equal to the frequency of
the atom, $\omega_{0}$, and the two side bands have frequencies
$\omega_{0}\pm\Omega$. That is, the interaction Hamiltonian
[\erf{eq.HamiltonianInteractionInt}] becomes
\begin{eqnarray} \label{eq.HamiltonianInteractionIntkk} \hat{V}_{\rm
int}(t)= i\hbar\sum_{k=-1}^{1}[\hat{\sigma}g^*_{k}e^{ik\Omega
t}\hat{a}_{k}\dg-\hat{\sigma}\dg g_{k}e^{-ik\Omega t}\hat{a}_{k}
].
\end{eqnarray}
With this interaction Hamiltonian the guiding state evolution is
determined by
\begin{eqnarray} \label{eq.IntSchEquationSimple2}
d_t\ket{{\Psi}(t)}&
=&\Big{\{}-\frac{i\Omega}{2}\hat{\sigma}_x+\sum_{k=-1}^{1}[\hat{\sigma}g^*_{k}e^{ik\Omega
t}\hat{a}_{k}\dg-\hat{\sigma}\dg g_{k}\nl\times e^{-ik\Omega
t}\hat{a}_{k} ]\Big{ \}} \ket{\Psi(t)}.
\end{eqnarray}
As in the last example, to calculate the stochastic evolution of
the property state
[$\ket{\Psi_{\{n_k(t)\}}(t)}=\ket{\{n_k(t)\}}\ket{\psi_{\{n_{k}\}}(t)}$],
we use Bell's solution to find
$J_{n_{-1},m_{-1};n_0,m_0;n_1,m_1}(t)$ [see Eq.(\ref{eq.BellJ})].
Here we find that only six values of
$J_{n_{-1},m_{-1};n_0,m_0;n_1,m_1}(t)$ are non-zero.

We expect that in the short time and large driving limit the
conditioned state will jump between approximate $\hat{\sigma}_x$
eigenstates ($\ket{{\sigma}_x=1}$ and $\ket{{\sigma}_x=-1}$). This
expectation can be illustrated analytically by considering the
following argument. We move to a second interaction frame, this
being the frame with both the free dynamics (standard interaction
frame) and $H_{\rm int}$ removed. That is, the interaction
Hamiltonian is
\begin{eqnarray}\label{eq.Vint3}
\hat{V}_{\rm
int'}(t)&=&i\hbar\sum_{k=-1}^{1}\Big{\{}g_k^{*}[\hat{\sigma}_x+\hat{\sigma}_x^{+}e^{i\Omega
t}+\hat{\sigma}_x^{-}\nl\times e^{-i\Omega t}]\hat{a}_k \dg
e^{ik\Omega t} -g_k[\hat{\sigma}_x+\hat{\sigma}_x^{+}e^{i\Omega
t}\nl+\hat{\sigma}_x^{-}e^{-i\Omega t}]\hat{a}_ke^{-ik\Omega
t}\Big{\}}/2,
\end{eqnarray} where
\begin{eqnarray}
\hat{\sigma}_x^{+}&=&\ket{{\sigma}_x=1}\bra{{\sigma}_x=-1}, \\
\hat{\sigma}_x^{-}&=&\ket{{\sigma}_x=-1}\bra{{\sigma}_x=1},
\end{eqnarray} and the prime denotes rotation to a second interaction frame.

We can now make the assumption that that for large driving all
sinusoidally varying terms can be neglected (this is effectively a
second RWA). Thus \erf{eq.Vint3} becomes
\begin{eqnarray}\label{eq.Vint4}
\hat{V}_{\rm int'}(t)&=&i\hbar\Big{\{}g_{-1}^{*}\hat{\sigma}_x^{+}
\hat{a}_{-1}\dg - g_{-1}\hat{\sigma}_x^{-}\hat{a}_{-1} +
g_0^{*}\hat{\sigma}_x \hat{a}_0 \dg \nl- g_0\hat{\sigma}_x
\hat{a}_0 +g_1^{*}\hat{\sigma}_x^{-}\hat{a}_1 \dg -
g_1\hat{\sigma}_x^{+}\hat{a}_1 \Big{\}}/2.\nl
\end{eqnarray} Moving back to the original interaction frame results in
\begin{eqnarray}\label{eq.Vint5}
\hat{V}_{\rm int}(t)&=&i\hbar\Big{\{}g_{-1}^{*}\hat{\sigma}_x^{+}
\hat{a}_{-1}\dg e^{-i\Omega t} -
g_{-1}\hat{\sigma}_x^{-}\hat{a}_{-1}e^{i\Omega t} + \nl
g_0^{*}\hat{\sigma}_x \hat{a}_0 \dg - g_0\hat{\sigma}_x \hat{a}_0
+g_1^{*}\hat{\sigma}_x^{-}\hat{a}_1 \dg e^{i\Omega t} \nl-
g_1\hat{\sigma}_x^{+}\hat{a}_1 e^{-i\Omega t}\Big{\}}/2.
\end{eqnarray} That is, in the large driving limit \erf{eq.HamiltonianInteractionIntkk} approximates
to the above Hamiltonian. Now since $\hat{\sigma}_x^{-}$ and
$\hat{\sigma}_x^{+}$ are lowering and raising operators for the
$\hat{\sigma}_x$ basis we can conclude that when there is a
transition from the $\ket{{\sigma}_x=1}$ to $\ket{{\sigma}_x=-1}$
eigenstates there is a corresponding upward jump in the photon
number for the higher-frequency mode or a lowering in the photon
number for the lower-frequency mode. The transition from the
$\ket{{\sigma}_x=-1}$ to $\ket{{\sigma}_x=1}$ eigenstate result in
jumps of opposite nature. A jump in the photon number for the
central-mode should not change the conditioned state.

\begin{figure}
\includegraphics[width= .45\textwidth]{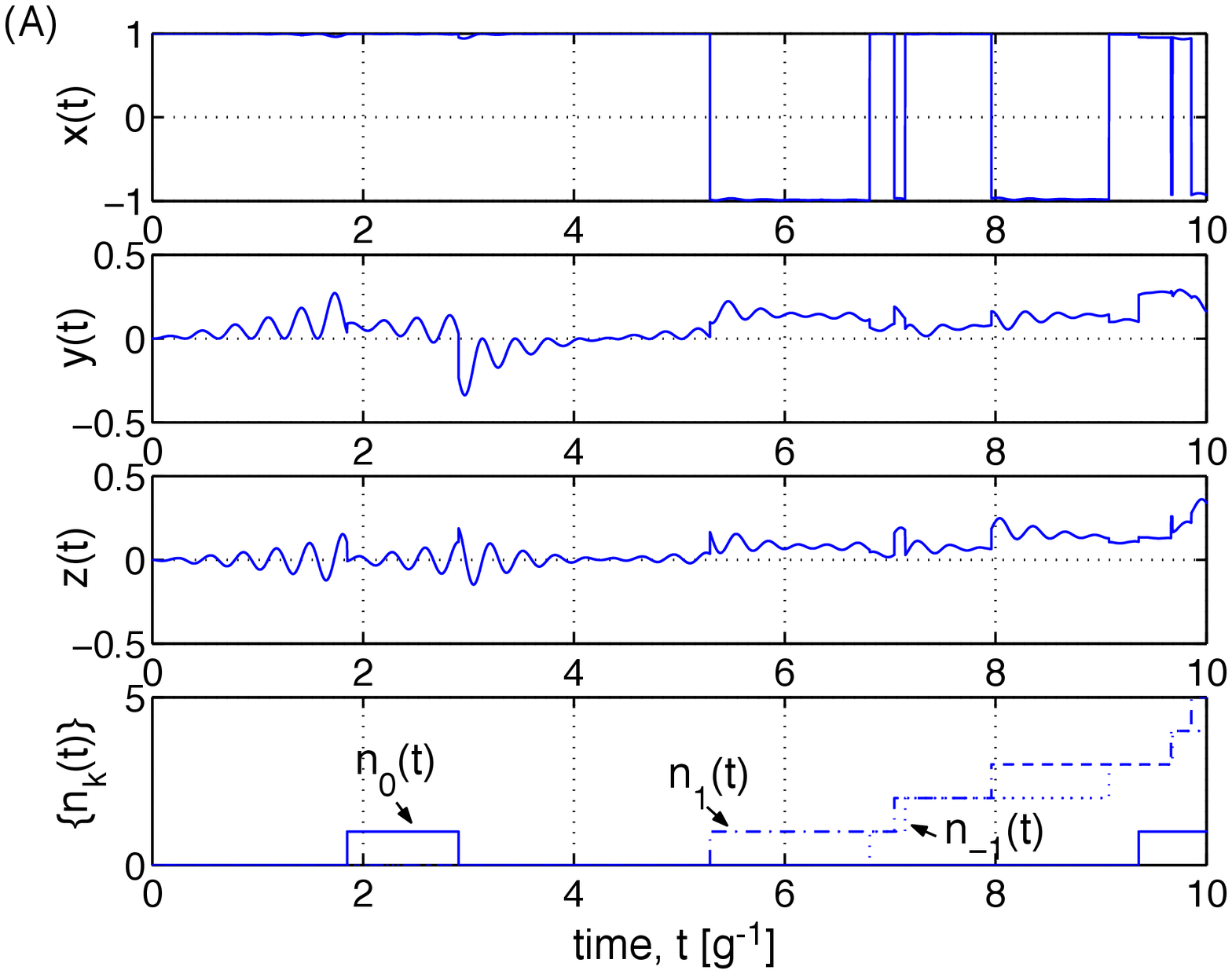}
\includegraphics[width= .45\textwidth]{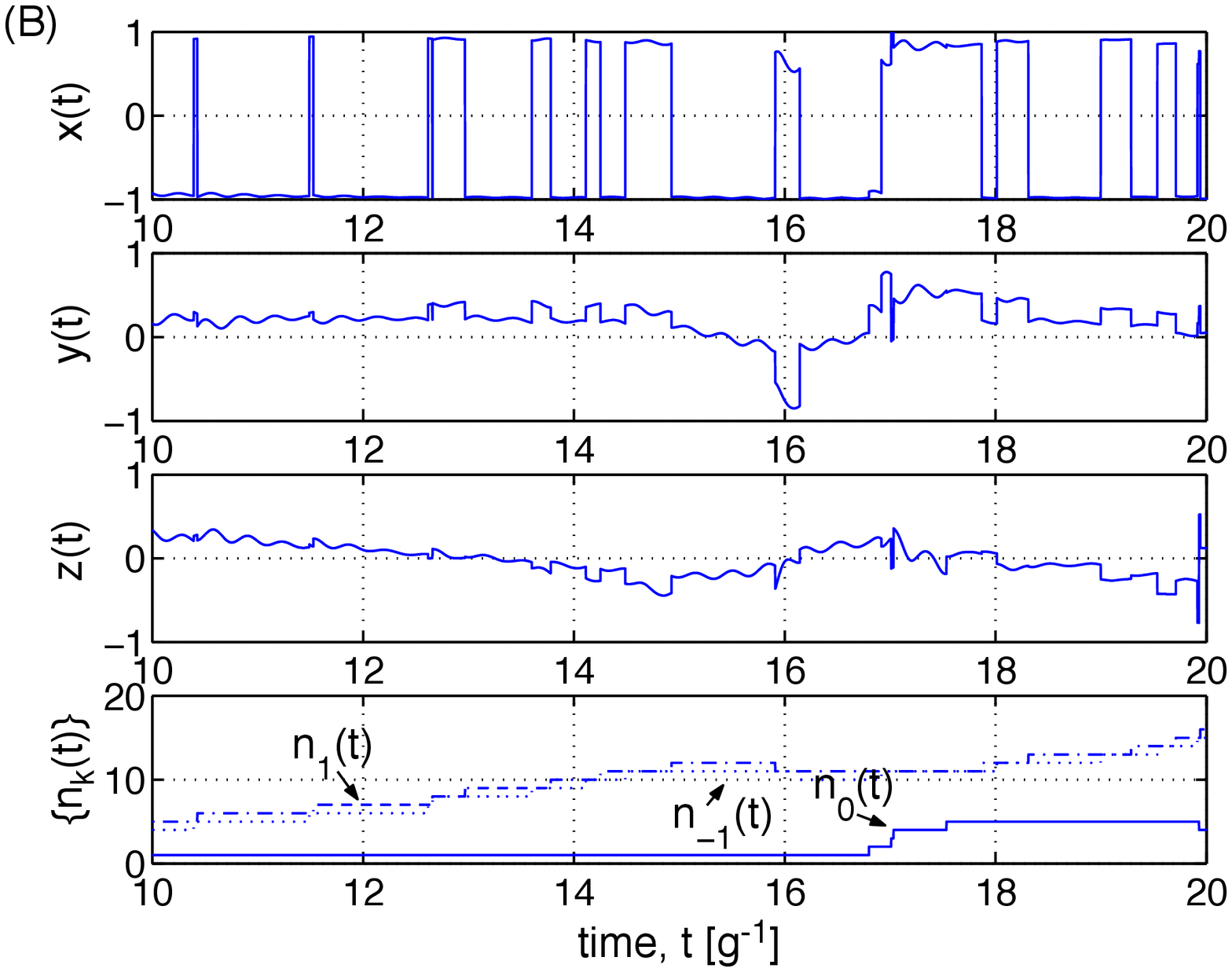}
\caption{\label{Fig.3ModeNumber} This figure depicts a typical
trajectory in Bloch representation for the spectral-mode
unraveling when the system is a driven TLA and the bath consists
of 3 modes. In this figure all calculations were done using the
initial system state $\ket{\psi(0)}=(\ket{e}+\ket{b})/\rt{2}$ with
parameters $g_{-1}=g_0=g_1=1$, and $\Omega=20g$. Time is measured
in units $g^{-1}$. Part (A) is for $0\leq t<10$ and part (B) for
$10\leq t<20$.}
\end{figure}

These predictions are verified in Figs. \ref{Fig.3ModeNumber} (A)
and (B) where we show simulations of the exact Bell dynamics with
interaction Hamiltonian \erf{eq.HamiltonianInteractionIntkk}. Note
the slight variation from the above prediction which we believe is
due to the fact that the driving is not infinite. The first figure
shows a typical trajectory for the conditioned state for
parameters $g_{-1}=g_0=g_1=1$, and $\Omega=20g$ for the first 10
units of time whereas the second figure shows the next 10 units of
time. It is in this second period of time where downward jumps in
the outer modes are observed.

\begin{figure}
\includegraphics[width= .45\textwidth]{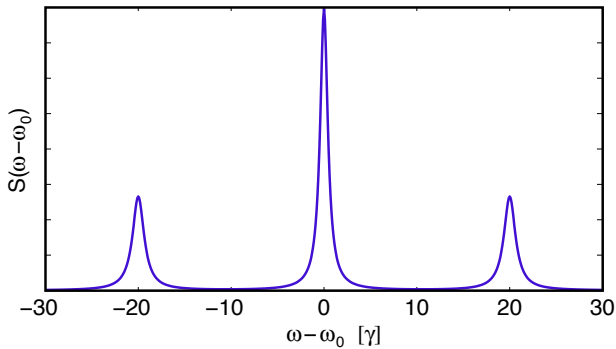}
\caption{\label{fig.Moll} This figure depicts the fluorescence
spectrum for a Markovian driven TLA with $\Omega$ equal to $20$
times the spontaneous emission rate, $\gamma$.}
\end{figure}

The prime motivation behind considering this 3-mode system is that
in the strong driving limit the jump dynamics should exhibit
similar features to those observed in a spectral-mode unraveling
of a Markovian open quantum system. This is because in the large
driving limit the fluorescence of a driving TLA takes the form of
a Mollow spectrum \cite{Mol75,Lou83} [Fig. \ref{fig.Moll} shows
this fluorescence spectrum (line graph) for $\Omega$ equal to $20$
times the spontaneous emission rate, $\gamma$]. In the
large driving limit the Mollow spectrum approaches becoming delta
functions centered on the frequencies $\omega_0-\Omega$,
$\omega_0$, and $\omega_0+\Omega$. Thus one would expect that in
this limit an unraveling in terms of spectral modes would have
similar dynamics in both a non-Markovian and Markovian open
quantum system. But what is the spectral-mode unraveling for a
Markovian open quantum system. In Ref. \cite{WisToo99} an
unraveling of a driven TLA with a Markovian system-bath
interaction using filter cavities and photdetectors was
introduced. We call this the Wiseman and Toombes filtered
Markovian (WTFM) unraveling. (Note this unraveling is an example
of an unraveling of a Markovian open quantum system which has
non-Markovian conditioned state evolution for the atom). In this
unraveling the field emitted by the atom is detected using two
cavities tuned to the sidebands of the Mollow triplet (these act
as mode filters). Thus in the strong driving limit the WTFM
unraveling effectively corresponds to a spectral-three-mode
unraveling. In Ref. \cite{WisToo99} it was observed that in this
limit the conditioned state for the TLA jumps between almost
$\hat\sigma_x$ eigenstates. Note, the non-Markovian case differs
from the WTFM in that it is possible to get downward jumps in
photon number. In the Markovian case this is not possible.
However, as before, these anomalous events become more frequent as
the modes become more populated.

\section{Temporal-mode unraveling}\label{Sec.XXX}

In the Markovian limit the simplest Markovian SSE is direct
detection. This detection scheme involves measurements performed
into temporal modes of the system (rather than the frequency
modes) \cite{Car93}. A temporal mode is best defined by
considering the electromagnetic field operator $\hat{\bf E}({\bf
r})$. The standard definition of this (neglecting polarization) is
\cite{TitGla66}
\begin{equation}
\hat{\bf E}({\bf r})=-i\sum_k^{\kappa}\hat{a}_k\dg {\bf
u}^{*}_k({\bf r})+i\sum_k^{\kappa}\hat{a}_k {\bf u}_k({\bf r})
\end{equation} where ${\bf u}_k({\bf r})$ if the mode function for
the $k^{\rm th}$ frequency (spectral) mode. This can be rewritten
as
\begin{equation}
\hat{\bf E}({\bf r})=-i\sum_\tau^{\kappa}\hat{b}_\tau\dg {\bf
v}^{*}_\tau({\bf r})+i\sum_\tau^{\kappa}\hat{b}_\tau {\bf
v}_\tau({\bf r}),
\end{equation}
where  ${\bf v}_\tau({\bf r})$ labels a new type of mode with
annihilation and creation operators $\hat{b}_\tau$ and
$\hat{b}_\tau\dg$. These new modes can be related to the frequency
modes by
\begin{equation}
  \hat{b}_{\tau}=\sum_{k}^{\kappa}\hat{a}_{k}\gamma_{\tau,k}^{*},
\end{equation} where $\gamma_{\tau,k}$ are the elements of a
unitary matrix
($\sum_{\tau}\gamma_{\tau,k}^{*}\gamma_{\tau,k'}=\delta_{k,k'}$ ).
Choosing $\gamma_{\tau,k}$ to give a discrete fourier transform,
\begin{eqnarray}\label{btau}
  \hat{b}_\tau&=&\frac{1}{\rt\kappa}\sum_{k}^{\kappa}\hat{a}_{k}\exp(-i2\pi \tau
  k/\kappa),\\
  \hat{a}_k&=&\frac{1}{\rt\kappa}\sum_{\tau}^{\kappa}\hat{b}_{\tau}\exp(i2\pi \tau
  k/\kappa),
\end{eqnarray} results in ${\bf v}_\tau({\bf r})$ having the functional form of a
temporal mode.

Since we have now defined an annihilation operator for the
temporal mode it is possible to define the observable to which we
wish to give an objective reality. This is the
temporal-mode-number operator,
$\hat{I}_{\tau}=\hat{b}_\tau\dg\hat{b}_\tau$. Thus the preferred
projective measure for this unraveling is
\begin{equation}
   \{\hat{\pi}_{n}\}= \{\hat{1}_{\rm sys}\otimes\ket{\{{I_{n_\tau}}\}}\bra{\{{I_{n_\tau}}\}}\},
\end{equation} where
$\ket{\{{I_{n_\tau}}\}}=\ket{I_{n_1}}\otimes...\otimes\ket{I_{n_\tau}}\otimes
...\otimes\ket{I_{n_\kappa}}$, and $\ket{I_{n_\tau}}$ is an
eigenstate of $\hat{I}_\tau$. This set of operators
$\{\hat{I}_\tau\}$ are the hidden variables for this unraveling.
Each operator $\hat{I}_\tau$ takes one of its possible integer
values $\{{I}_{n_\tau}\}$.

To provide a clearer picture of this unraveling we consider
briefly the Markovian case. In the Markovian limit the number of
modes becomes continuous ($\kappa\rightarrow\infty$) and the
system-bath coupling becomes flat ($g_k\rightarrow\rt{\gamma/2}$).
As such we must define continuous (in frequency) annihilation
operators $\hat a(\tilde{\omega})$ where
$\tilde{\omega}=\omega-\omega_0$ with $\omega$ being a continues
variable labelling the frequency of the bath. This results in the
temporal modes, denoted by $\hat{b}(\tau)$, becoming continuous in
time and being related to $\hat a(\tilde{\omega})$ by the fourier
transform
\begin{eqnarray}
  \hat{b}(\tau)&=&\frac{1}{\rt{2\pi}
  }\int_{-\infty}^{\infty}\hat{a}(\tilde\omega)\exp(-i\tilde\omega
  \tau) d\tilde\omega., \\
 \hat{a}(\tilde\omega)&=&\frac{1}{\rt{2\pi}
  }\int_{-\infty}^{\infty}\hat{b}(\tau)\exp(i\tilde\omega
  \tau) d\tau
\end{eqnarray}
Furthermore the interaction Hamiltonian,
\erf{eq.HamiltonianInteractionInt} (this is in the interaction
frame with the free system and bath dynamics removed) becomes
\begin{equation} \label{eq.HamiltonianInteractionIntMar2}
\hat{V}_{\rm int}(t)=
i\hbar\rt{\frac{\gamma}{2\pi}}\int_{-\omega_0}^{\infty}[\hat{\sigma}e^{i\tilde{\omega}
t}\hat{a}\dg(\tilde{\omega})-\hat{\sigma}\dg e^{-i\tilde\omega
t}\hat{a}(\tilde\omega) ]d\tilde\omega.
\end{equation}
Assuming that $\omega_0$ is large (this is valid for optical
situations), with little error the lower limits of the integrals
can be taken as $\infty$ \cite{GarZol00}. This then allows us to
rewrite \erf{eq.HamiltonianInteractionIntMar2} as
\begin{equation} \label{eq.HamiltonianInteractionIntMar3}
\hat{V}_{\rm int}(t)=
i\hbar\rt{{\gamma}}\int_{-\infty}^{\infty}[\hat{\sigma}\delta(\tau-t)\hat{b}\dg(\tau)-\hat{\sigma}\dg
\delta(t-\tau)\hat{b}(\tau) ]d\tau.
\end{equation}
Here we see that at time $t$ only one temporal mode is coupled to
the system. This is precisely why the bath in a Markovian open
quantum system only effects the system at one time (no memory
effects).

For a non-Markovian open quantum system the above delta
correlations will not exists and as such the bath will have a
memory. Writing \erf{eq.HamiltonianInteractionInt} in terms of the
discrete temporal modes [\erf{btau}] gives
\begin{eqnarray} \label{eq.HamiltonianInteractionIntSimple3} \hat{V}_{\rm
int}(t)&=&i\hbar\sum_{\tau}^{\kappa}[c_\tau^{*}(t)\hat\sigma\hat{b}_\tau\dg-c_\tau(t)\hat\sigma\dg\hat{b}_\tau],
\end{eqnarray} where
\begin{eqnarray}
c_\tau(t)&=&\frac{1}{\rt{\kappa}}\sum_{k}^{\kappa} g_k
e^{-i\Omega_k t+i2\pi \tau
  k/\kappa}.
\end{eqnarray} This clearly does not have a delta correlation
between $t$ and $\tau$ as occurs in the Markovian case
[\erf{eq.HamiltonianInteractionIntMar3}]. To show that this does
occur in the Markovian limit we simply let $g_k$ become flat
($g_k=g$) and assume a constant spacing in frequency $\Omega_k=k
\Omega$. Doing this results in
\begin{equation}
c_\tau(t)=\frac{g}{\rt{\kappa}}\Big{\{}1+2\sum_{k=1}^{(\kappa-1)/2}
\cos[k(\Omega t-2\pi \tau /\kappa)]\Big{\}}.
\end{equation} This by definition is a Kronecker $\delta$-function in the $\kappa\rightarrow\infty$
limit. This in turn implies a Dirac $\delta$-function in the
continuous limit.

\begin{figure}
\includegraphics[width= .45\textwidth]{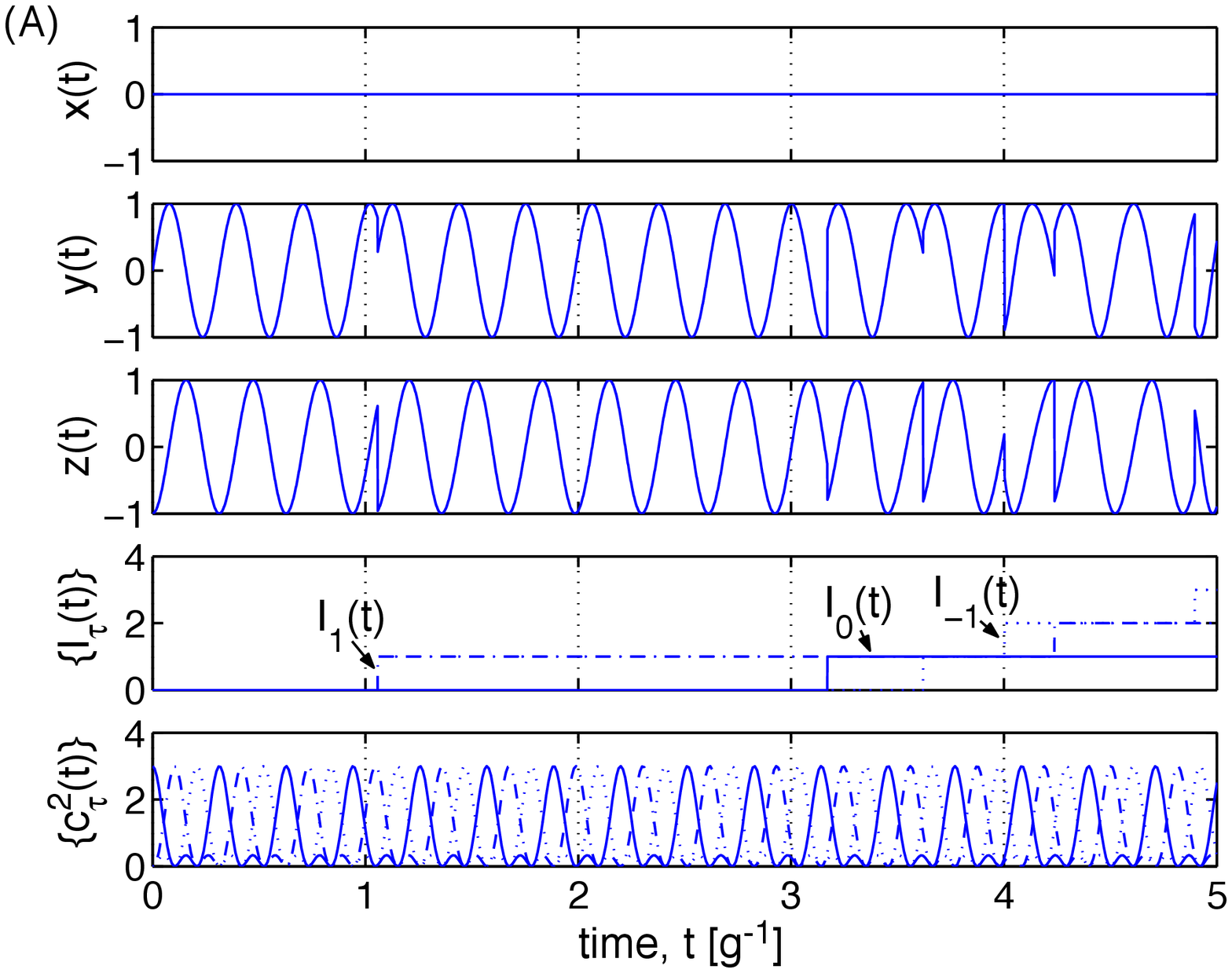}
\includegraphics[width= .45\textwidth]{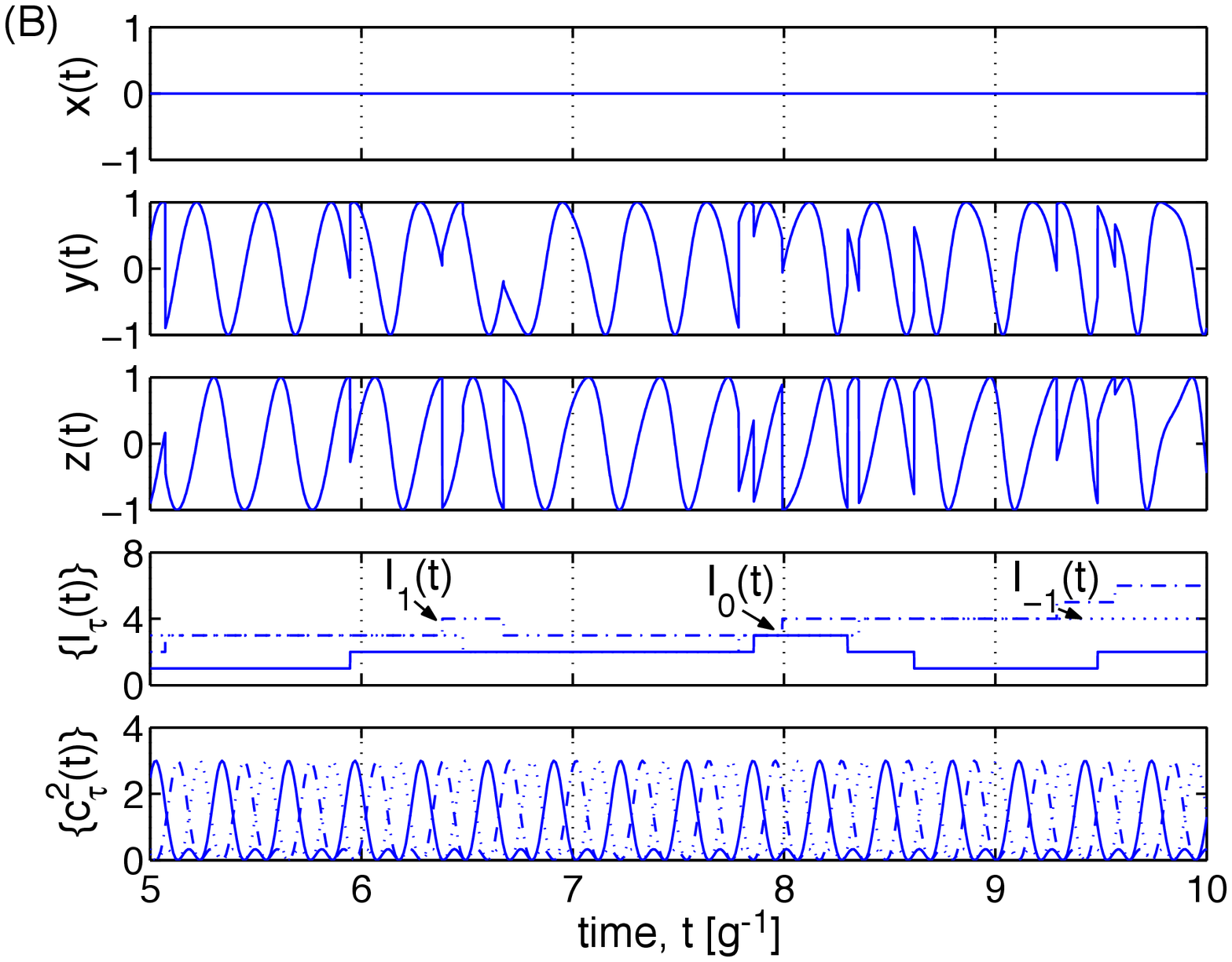}
\caption{\label{Fig.3Modeb} This figure depicts a typical
trajectory in Bloch representation for the temporal-mode
unraveling when the system is a driven TLA and the bath consists
of 3 modes. In this figure all calculations were done using the
initial system state $\ket{\psi(0)}=\ket{b}$ with parameters
$g_{-1}=g_0=g_1=1$, and $\Omega=20g$. Time is measured in units
$g^{-1}$. Part (A) is for $0\leq t<5$ and part (B) for $5\leq
t<10$.}
\end{figure}

To illustrate this unraveling we consider a driven TLA coupled to
a three mode bath as before. For the three mode case \erf{btau}
becomes
\begin{equation}\label{btau2}
  \hat{b}_\tau=\frac{1}{\rt{3}}\sum_{k=-1}^{1}\hat{a}_{k}\exp(-i2\pi \tau
  k/3).
\end{equation}
Using these operators we can rewrite the interaction Hamiltonian,
\erf{eq.HamiltonianInteractionIntSimple3}, as
\begin{eqnarray} \label{eq.HamiltonianInteractionIntSimple4} \hat{V}_{\rm
int}(t)&=&i\hbar
\sum_{\tau=-1}^{1}[c_\tau(t)\hat\sigma\hat{b}_\tau\dg-c_\tau(t)\hat\sigma\dg\hat{b}_\tau].
\end{eqnarray} where
\begin{eqnarray}
c_\tau(t)=\frac{g}{\rt{3}}\Big{\{}1+ 2 \cos(\Omega t-2\pi \tau
/3)\Big{\}}.
\end{eqnarray}
Here we have assumed $g_{-1}=g_0=g_1=g$ as before. Although these
functions are not $\delta$-functions, $c_\tau^2(t)$ are peaked at
times $t=2\pi(\tau/3+n)/\Omega$ for $n$ an integer, as shown in
the finial plot of Fig. \ref{Fig.3Modeb}.

Using the same techniques as in the last unraveling it is possible
to determined a probability current
$J_{n_{1},m_{1};n_2,m_2;n_3,m_3}(t)$. Then we can find a typical
trajectory for the values of the bath hidden variables
$\{I_{n_\tau}(t)\}$ and a numerical trajectory for the condition
state $\ket{\psi_{\{I_{n_\tau}(t)\}}(t)}$. Figures
\ref{Fig.3Modeb} (A) and (B) illustrates a typical trajectory for
a driven TLA with driving frequency $\Omega=20g$. Here we see that
the dynamics are clearly unlike that observed for a Markovian open
quantum system under a temporal-mode unraveling (direct
detection). In the Markovian case an upward jump in the temporal
mode corresponds to the conditioned system state jumping to the
ground state ($z=-1$); see Ref. \cite{Car93}. However for an
upward jump into the $\tau^{\rm th} $ temporal mode which occurs
when $c_\tau^2(t)$ is a maximum this behaviour is observed, and
the closer $c_\tau^2(t)$ is to maximum the closer the conditioned
state jumps to $z=-1$. Thus we expect that when there are enough
modes such that at every time $t$ one of the $c_\tau^2(t)'s$ is a
maximum (Markovian limit) an upward jump in photon number will
result in the system state jumping to the ground state.  Note, as
in all the other examples at later times when the modes become
populated extra non-Markovian features are present. Firstly
downward jumps in temporal mode number are present. These result
in the conditioned state jumping towards the excited state
($z=1$). But more interesting at approximately $t=4.9g^{-1}$ (as
well as at $t=9.5g^{-1}$), there is a upward jump in the $\tau=-1$
($\tau=0$) temporal mode which results in an increasing of atomic
energy (the conditioned state jumping towards a higher atomic
energy state).

\section{Comparison with Jack, Collet, and Walls}\label{Sec.Coll}

Recently Jack, Collet and Walls (JCW) have published a series of
papers, Refs. \cite{JacColWal99,JacColWal99b,JacCol00}, where they
developed a quantum trajectory theory (continuous in time
measurement theory) for non-Markovian systems which incorporates
jump-like unravelings. Thus the aim of this section is to review
their theory and discuss how it differs from ours.

In the JCW theory, because it concerns measurements, they have to
add the requirement, ``a measurement at a certain time does not
disturb any future measurements" \cite{JacCol00}. This basically
means that the average state of the system (average over all
possible measurements records) must be equal to the reduced state.
Thus there is no back action on the system from the measurements.
How is this possible? For argument's sake, consider two
consecutive quantum non-demolition measurements of the bath
observable
\begin{equation}
  Z=\{(z_n,\hat{1}_{\rm sys}\otimes\ket{z_n}\bra{z_n})\}.
\end{equation} Then we can write the average state of the system
as
\begin{eqnarray}
  \rho(t)&=&\sum_{n,m}\bra{z^{(2)}_n}\hat{U}(t_2,t_1)\ket{z^{(1)}_m}\bra{z^{(1)}_m}\hat{U}(t_1,0)\ket{\Psi(0)}
  \nl\times
  \bra{\Psi(0)}\hat{U}\dg(t_1,0)\ket{z^{(1)}_m}\bra{z^{(1)}_m}\hat{U}\dg(t_2,t_1)\ket{z^{(2)}_n}\nn\\
\end{eqnarray}
where the superscripts $(1)$ and $(2)$ refer to the first
measurement made at time $t_1$ and the second measurement made at
time $t_2$. This is only equal to the reduced state,
\erf{eq.RhoFromTrace}, if
$[\hat{U}(t_2,t_1),\ket{z^{(1)}_m}\bra{z^{(1)}_m}]=0$. This
implies that either $\hat{U}(t_2,t_1)$ is diagonal in the bath
basis $\{\ket{z^{(1)}_m}\}$ or $\hat{U}(t_2,t_1)$
 and $\{\ket{z^{(1)}_m}\bra{z^{(1)}_m}\}$ operate on
separate Hilbert spaces. This is precisely what happens in the
Markovian limit (because of the delta correlations between the
bath noise operators \cite{GarZol00}). JCW theory uses the later
case also and proposes that we can define a field quantity
$\hat{\phi}(t)$ to be measured, such that
\begin{equation}
  [\hat{\phi}(t),\hat{\phi}\dg(t')]=\delta(t-t').
\end{equation} However, apart from Markovian open quantum systems it is not clear what measurement
schemes this is applicable too. JCW consider in Ref.
\cite{JacColWal99} a non-Markovian unraveling of a Markovian open
quantum system, this being their spectral detection unraveling.
This is similar to the WTFM unraveling, except they do not extend
the system to include cavities.

In the JCW theory to account for the the memory effects of
non-Markovian unravelings it is imposable to assign a pure state
for the system at time $t$ conditioned on a past measurement
record, ${\bf I}_{[0,t]}$ (a string of results $r_1,..., r_k$
where $t=k\delta t$). Instead the state of the system conditioned
on the measurement record ${\bf I}_{[0,t]}$ is given by
\begin{equation}
  \rho_{{\bf I}_{[0,t]}}(t)=\frac{\tilde\rho_{{\bf I}_{[0,t]}}(t)}{{\rm Pr}({\bf
  I}_{[0,t]})},
\end{equation}
where
\begin{equation}
  {\rm Pr}({\bf I}_{[0,t]})=\int d \{r_{\frac{s}{\delta t}>\frac{t}{\delta t}}\}
   \bra{\tilde\psi_{{\bf I}_{[0,s]}}(t)} {\tilde\psi_{{\bf
   I}_{[0,s]}}(t)}\rangle
\end{equation}
and
\begin{equation}
 \tilde\rho_{{\bf I}_{[0,t]}}(t)=\int d \{r_{\frac{s}{\delta t}>\frac{t}{\delta t}}\}
 \ket{\tilde\psi_{{\bf I}_{[0,s]}}(t)}\bra{\tilde\psi_{{\bf I}_{[0,s]}}t)}.
\end{equation}
Here $\ket{\tilde\psi_{{\bf I}_{[0,s]}}(t)}$ is the the state of
the system at time $t$ conditioned on the complete record for all
time (measurement which have not yet come to be). In a sense this
is like a retrodictive state and may have some interpretation
under retrodictive quantum mechanics \cite{PEGG}. Thus we see that
the best we can do is to assign a mixed state to the system, this
being a mixture over all possible future records.

However, as JCW point out, if we can assume a finite-time memory
function (a memory function which is zero for time less then
$t-t_m$ in the past), then we can assign a pure state to the
system at time $t-t_m$ conditioned on measurements up and until
time $t$. That is, the state of the system conditioned on the
record ${\bf I}_{[0,t]}$  is $\ket{\tilde\psi_{{\bf
I}_{[0,t]}}(t-t_m)} $ (when appropriately normalized).

To summarize, while their theory seems to be correct we are not
sure of the applicability of it. In general non-Markovian systems
will have back action effects if a measurement is performed, so
the average state for consecutive measurements will not be the
reduced state. Also the pure states they associate for the system
are not the same as our conditioned states. In fact in Ref.
\cite{GamWis002} we came to the conclusion that the only possible
interpretation of non-Markovian SSEs (our conditioned states)
under the orthodox theory is they are numerically tools used to
generate the correct state of the system at time $t$ given that a
measurement has been performed on the bath at this time yielding
the appropriate result.

\section{Discussion and conclusion}\label{Sec.Conclude}

In this paper we have investigated non-Markovian unravelings that
exhibit jump-like behaviour. We observed that it is impossible to
define an analytical non-Markovian SSE for these unravelings, but
by using the modal interpretation of quantum mechanics, namely
Bell's beable dynamics \cite{Bel84}, it is possible to calculate
the numerical equivalent to the solution of a non-Markovian SSE.
To illustrate this we considered an open quantum system consisting
of a TLA coupled to a bath of harmonic oscillators.

The first example we considered was a driven two level atom (TLA),
driven with Rabi frequency $\Omega$, and coupled linearly to a
single mode bath. Here we observed that the non-Markovian dynamics
(conditioned states evolution) are quite different in nature to
Markovian dynamics. For example we observe that it possible to get
upward jumps in photon number which result in an increasing of the
atomic energy of the atom.

The second example was again a driven TLA, but we considered the
bath to contain three modes. The central mode of the bath had a
frequency equal to the frequency of the atom, $\omega_{0}$, while
the two outer modes had frequencies $\omega_{0}\pm\Omega$. This
was chosen as the spectrum of this system (three modes) is similar
to the fluorescence spectrum of a driven TLA (Mollow spectrum
\cite{Mol75}) in the strong driving and Markovian limit (see Fig.
\ref{fig.Moll}). Thus one expects that the conditioned states
evolution should contain some features which are consistent with
the dynamics of a Markovian open quantum system. To illustrate
this we considered the two jump-like unravelings: the spectral
mode and temporal-mode. It was observed that in both these
unravelings the jump-like dynamics did contain some features of
the equivalent Markovian open quantum system.

In conclusion, by accepting the modal interpretation of
non-Markovian SSEs it is possible to generalize non-Markovian
unravelings to include jump-like unravelings, but it is only
possible to numerically determine the solution. Note this
generalization also applies to diffusive non-Markovian SSEs. That
is, we can extend the coherent
\cite{DioStr97,DioGisStr98,StrDioGis99,GamWis002,GamWis006},
position \cite{GamWis005,GamWis006} and quadrature
\cite{BasGhi02,GamWis002,GamWis005} unraveling to include all
possible choices of a continuous preferred projective (and
positive operator) measures. However, in the diffusive case it
maybe possible to derive a generalized analytical non-Markovian
SSE. Bassi \cite{Bas03} has already proceeded down this path by
calculating a generalized linear non-Markovian SSEs, for diffusive
unravelings, but further question still remain. For example, what
subset of reduced states do these linear non-Markovian SSEs belong
to and how can an extension to the normalized case be made?
Furthermore although it maybe possible to write an analytical
expression for a diffusive non-Markovian SSE we believe that, in
general, due to the functional derivative the solution of this
equation will only be determined by numerical perturbative
techniques (see \cite{YuDioGisStr99} or
 \cite{GamWis003} for perturbative approximations) which in
some circumstances effectively amounts to solving the \sch
equation for the total state $\ket{\Psi(t)}$.

\begin{acknowledgments}
This work was supported by the Australian Research Council. T.A.
was supported by the University of Stockholm. J.G. acknowledges
the use of the Queensland Parallel Supercomputing Facility.
\end{acknowledgments}


\begin{thebibliography}{}

\bibitem{DioStr97}
L. Di\'osi and W.T. Strunz, Phys. Lett. A {\bf 235}, 569 (1997).


\bibitem{DioGisStr98}
L. Di\'osi, N. Gisin, and W.T. Strunz, Phys. Rev. A {\bf  58},
1699 (1998).


\bibitem{StrDioGis99}
W.T. Strunz, L. Di\'osi, and N. Gisin, Phys. Rev. Lett. {\bf 82},
1801 (1999).


\bibitem{YuDioGisStr99}
T. Yu, L. Di\'osi, N. Gisin, and W.T. Strunz, Phys. Rev. A {\bf
60}, 91 (1999).


\bibitem{Cre00}
J. D. Cresser, Laser Phys. {\bf  10}, 1 (2000).

\bibitem{Bud00}
A. A. Budini, Phys. Rev. A {\bf 63}, 012106 (2000).

\bibitem{BasGhi02}
A. Bassi and G. C. Ghirardi, Phys. Rev. A {\bf 65}, 042114 (2002).


\bibitem{GamWis002}
J. Gambetta and H. M. Wiseman, Phys. Rev. A {\bf 66}, 012108
(2002).

\bibitem{GamWis003}
J. Gambetta and H. M. Wiseman, Phys. Rev. A {\bf 66}, 052105
(2002).

\bibitem{Bas03}
A. Bassi, Phys. Rev. A {\bf 67}, 062101 (2003).


\bibitem{GamWis005}
J. Gambetta and H. M. Wiseman, Phys. Rev. A {\bf 68}, 062104
(2003).

\bibitem{GamWis006}
J. Gambetta and H. M. Wiseman, to be published in J. Opt B.

\bibitem{Car93}
H. J. Carmichael, {\em An Open System Approach to Quantum Optics}
(Springer, Berlin, 1993).

\bibitem{Fra81}
B. van Fraassen, in {\em Current Issues in Quantum Logic}, edited
by E. Beltrametti and B. van Fraassen (World Scientific,
Singapore, 1981), pp. 229-258.

\bibitem{Hea89}
R. Healy, {\em The Philosophy of Quantum Mechanics}  (Cambridge
University Press, Cambridge, 1989).

\bibitem{Die97}
D. Dieks, in {\em Quantum Measurements: Beyond Paradox}, edited by
R. A. Healey and G. Hellman (University of Minnesota Press,
Minneapolis, 1997), pp. 144-159.

\bibitem{Bub97} J. Bub, {\em Interpretating the Quantum World} (Cambridge University Press, Cambridge, 1997).

\bibitem{BacDic99} G. Bacciagaluppi and M. Dickson, Found. Phys. {\bf 29}, 1165 (1999).

\bibitem{Sud00} A. Sudbery, Stud. Hist. Philos. Mod. Phys. {\bf 33}, 387 (2002).

\bibitem{GamWis004} J. Gambetta and H. M. Wiseman, Found. Phys. {\bf 34}, 419 (2004).

\bibitem{SpeSip01b}
R. W. Spekkens and J. E. Sipe, Found. Phys. {\bf 31}, 1431 (2001).

\bibitem{Bel84} J. S. Bell, CERN-TH.4035/84, (1984). Reprinted in
                {\em John S. Bell on the Foundations of Quantum Mechanics}, edited
                by M. Bell, K. Gottfried, and M. Veltman (World Scientific,
                Singapore, 2001).

\bibitem{Boh52a}
D. Bohm, Phys. Rev. {\bf 85}, 166 (1952).

\bibitem{BasGhi03}
A. Bassi and G. C. Ghirardi, Phys. Rep. {\bf 379}, 257 (2003).

\bibitem{Mol75}
B. R. Mollow, Phys. Rev. A {\bf 12}, 1919 (1975).

\bibitem{JacColWal99}
M.W. Jack, M.J. Collett, and D.F. Walls,  Phys. Rev. A {\bf 59},
2306 (1999).

\bibitem{JacColWal99b}
M.W. Jack, M.J. Collett, and D.F. Walls, J. Opt. B: Quantum
Semiclass. Opt. {\bf 1}, 452 (1999).

\bibitem{JacCol00}
M.W. Jack and M.J. Collett, Phys. Rev. A {\bf 61}, 062106 (2000).

\bibitem{BreKapPet99}
H. Breuer, B. Kappler, and F. Petruccione, Phys. Rev. A {\bf 59},
1633 (1999).

\bibitem{Bre03}
H. Breuer, quant-ph/0308052.

\bibitem{Vin93}
J.C. Vink, Phys. Rev. A {\bf 48}, 1808 (1993).

\bibitem{Dic97}
M. Dickson, in {\em Quantum Measurements: Beyond Paradox}, edited
by R. A. Healey and G. Hellman (University of Minnesota Press,
Minneapolis, 1997), pp. 160-182.

\bibitem{Lou83}
R. Loudon, {\em The Quantum Theory of Light} (Oxford University
Press, New York, 1983).

\bibitem{WisToo99}
H. M. Wiseman and G.E. Toombes,  Phys. Rev. A {\bf 60}, 2474
(1999).

\bibitem{GarZol00}
C. W. Gardiner, and P. Zoller, {\em Quantum Noise} (Springer,
Berlin, 2000).

\bibitem{TitGla66}
U. M. Titulaer and R. J. Glauber, Phys. Rev. {\bf 145}, 1041
(1966).

\bibitem{PEGG}
S. M. Barnett, D. T. Pegg, and J. Jeffers, J. Mod. Opt. {\bf 47},
1779 (2000).

\end{thebibliography}
\end{document}